\documentclass[a4paper,11pt]{article}
\usepackage{pos}
\usepackage{graphicx}
\usepackage{subcaption}
\usepackage[backend=biber,style=phys]{biblatex}

\AtEveryBibitem{\clearfield{title}}
\title{H.E.S.S. observations of SN 2024ggi}

\author[a]{J. Borowska-Naguszewska}\author*[b,c,d]{R. Brose}\author[e]{B. Cornejo} \author[d]{J. Mackey} \author[a]{R.~D.~Parsons} \author[e]{F. Sch\"ussler} \onbehalf{on behalf of the H.E.S.S. Collaboration}
\storeFirstAuthor{J. Borowska-Naguszewska}
\affiliation[a]{Humboldt-Universität zu Berlin, Faculty of Mathematics and Natural Sciences,
Newtonstraße 15, 12489 Berlin, Germany}
\affiliation[b]{Institute of Physics and Astronomy, University of Potsdam, 14476 Potsdam-Golm, Germany}
\affiliation[c]{School of Physical Sciences and Centre for Astrophysics \& Relativity, Dublin City University, Glasnevin, D09 W6Y4, Ireland}
\affiliation[d]{Dublin Institute for Advanced Studies, Astronomy \& Astrophysics Section, DIAS Dunsink Observatory, Dublin D15 XR2R, Ireland}
\affiliation[e]{IRFU, CEA, Universit\'e Paris-Saclay, F-91191 Gif-sur-Yvette, France}

\emailAdd{jowita.borowska@physik.hu-berlin.de}

\abstract{Supernova (SN) explosions interacting with dense circumstellar medium are considered to be very promising sites for efficient cosmic-ray (CR) acceleration and subsequent emission of neutral-pion-decay gamma rays. These environments share similarities with already detected gamma-ray novae, but with much greater available energy content, so it is important to characterize their emission in the very-high-energy range. We present the results of H.E.S.S. observations of one such candidate source -- SN 2024ggi, located in NGC 3621 at a distance of 7.24 Mpc. A total of 30 hours of data, gathered throughout a month of post-explosion observations, provide flux upper limits that are used to constrain source parameters, offering meaningful insights for theoretical predictions. We exclude bright gamma-ray emission in the first day after explosion, and later upper limits are consistent with wind densities derived from optical observations.}

\ConferenceLogo{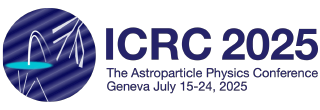}

\FullConference{39th International Cosmic Ray Conference (ICRC2025)\\
 15–24 July 2025\\
Geneva, Switzerland}

\begin{document}
\maketitle

\section{Introduction}

Core-collapse supernovae (SNe) represent the final phase in the evolution of massive stars. An increasing number of observations indicates that the explosion is often preceded by a several-year period of instability, during which the progenitor star experiences mass loss episodes \cite{Smith_2014, Strotjohann_2021}. The expelled material itself creates a shell of circumstellar medium (CSM), which the supernova ejecta interact with following the explosion. In the resulting shock, a significant part of their kinetic energy is then converted to photons that are radiated away \cite{Murase_2014_scheme, Smith_2017}. The emission is anticipated to span the whole electromagnetic spectrum. In the very-high-energy domain it is closely linked with the processes responsible for the acceleration of cosmic rays (CRs), that may reach energies even beyond the PeV regime \cite{Bykov_2018}. Accelerated particles interact with ions in the surrounding CSM, producing hadronic gamma-ray emission via neutral-pion decay, at the level potentially observable by current and future instruments \cite{Cristofari_2020, Cristofari_2022}. This is supported by the fact that conditions similar to those in interaction-powered SNe prevail in already gamma-ray detected systems, such as colliding wind binaries, like Eta Carinae \cite{Ohm_2015}, or
the recurrent nova RS Ophiuchi \cite{rsoph_hess}, where the material expelled by a white dwarf runs into the companion star's wind. While SN explosions are vastly more energetic than novae, they are yet to be seen in the highest energy range. 
Distance to the source seems to be the key parameter hindering the detection, but the signal is also severely suppressed when the TeV gamma rays emitted at the shock wave interact with the lower-energy photon field of the SN photosphere, producing electron-positron pairs ($\gamma\gamma$ absorption) \cite{Cristofari_2020, Cristofari_2022}. The exact impact of this effect depends on geometry and the temporal evolution of the SN, which are both subject to modeling uncertainties.

\subsection{H.E.S.S. core-collapse SN follow-up program}
The High Energy Stereoscopic System (H.E.S.S.) is an array of five Imaging Atmospheric Cherenkov Telescopes located in the Khomas Highland of Namibia at 1800 m above sea level. Four telescopes have the effective mirror diameter of 12 m (CT1--4) \cite{Aharonian_2006} and are arranged in a square formation with sides of 120 m. The fifth, 28-m telescope, (CT5) \cite{2015ICRC...34..847V} is placed in the center to observe fainter air showers, extending sensitivity towards lower energies. The observatory operates an ongoing follow-up program dedicated to core-collapse SN, which are usually first discovered by wide-field-of-view optical surveys, and then announced through relevant communication channels. The horizon of detectability by H.E.S.S. has been estimated to reach a distance of a few Mpc \cite{Brose_2022}, although this depends strongly on the combination of characteristics related to the stellar winds and progenitors (such as mass-loss rate or wind velocity). Some exceptionally bright sources might be detectable even if located much further away. That includes the class of extreme objects called Fast Blue Optical Transients (FBOTs), that can also exhibit signatures of the interaction mechanism in their spectra \cite{Fox_2019}, causing them to be classified as SNe. Their high luminosity can be associated with very high shock velocities, while fast rise times indicate thin, dense CSM shells \cite{Ho_2020} and a small mass of SN ejecta \cite{Lyutikov_2022}.
The level of expected gamma-ray signal is influenced by the two-photon annihilation process, so the observations should be conducted when the suppression effect is the weakest. The exact timescale varies between candidate sources, but it has been shown that the detection probability is highest either before 1 day or after 10 days following the explosion \cite{Cristofari_2020}. 

Due to a relatively low predicated rate of core-collapse SNe within the most relevant distance range (1-2 per year within 10 Mpc \cite{Smartt_2009}), only a limited number of dedicated H.E.S.S. observations have been performed to date. So far, the publications (finding no significant excess, but deriving valuable constraints) include nine nearby SNe, that have been serendipitously observed earlier than a year after their discovery (in the data gathered between the end of year 2003 and 2004), as well as the SN 2016adj at 3.8 Mpc \cite{2019}, and AT 2019krl at 9.8 Mpc \cite{HESS_ccSN}, both observed as a Target of Opportunity. Moreover, the fast-rising AT 2018cow at 63 Mpc was also observed, resulting in flux upper limits  \cite{atel11956}.

\section{Observations and data analysis}

\begin{figure}[b]
  \centering
  \includegraphics[width=0.8\textwidth]{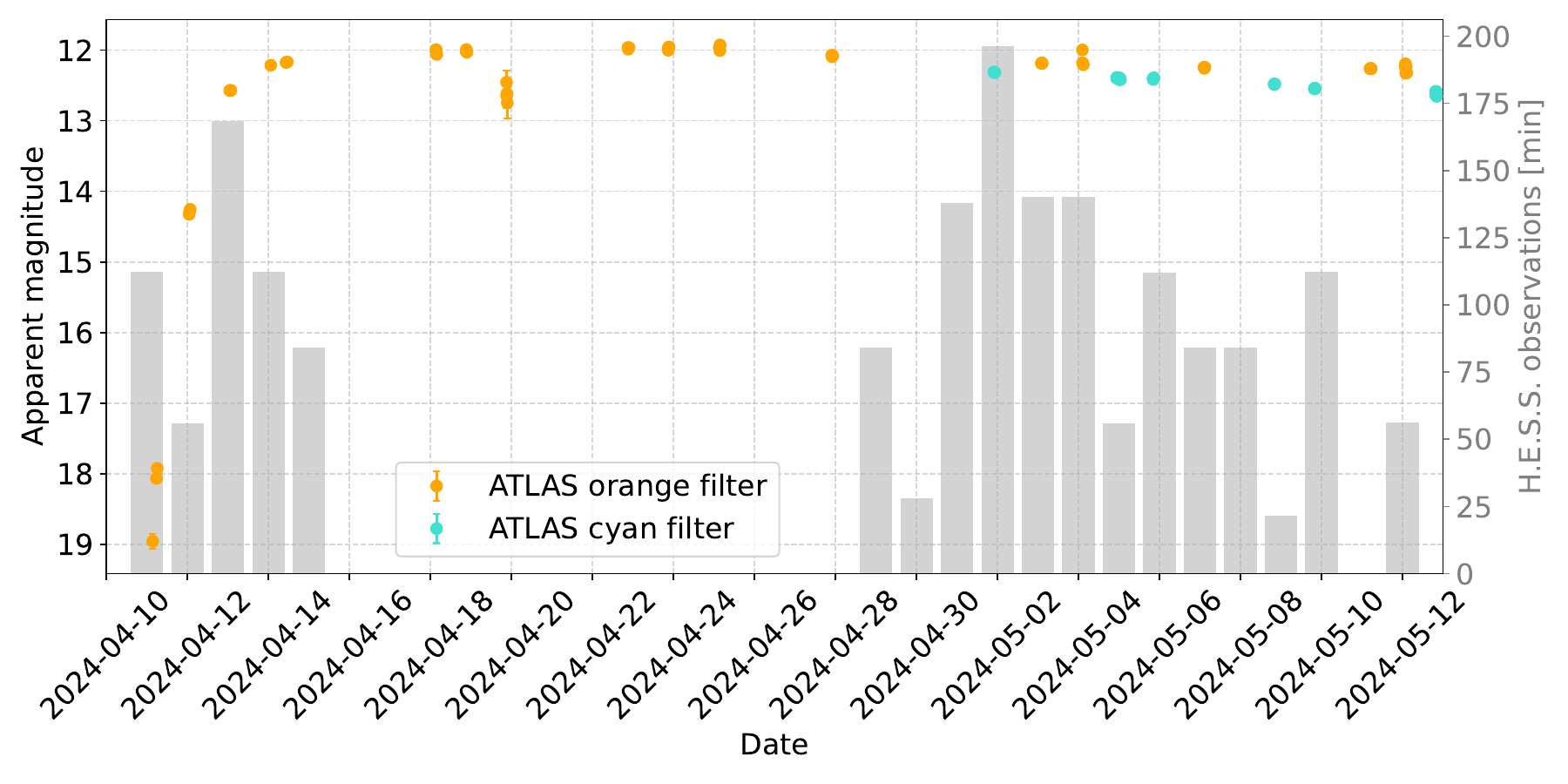}
  \caption{All selected H.E.S.S. observations of SN 2024ggi spanning the whole observation period, together with the optical lightcurve from the ATLAS forced photometry server \cite{Tonry_2018,atlas_transients,forcedphotoserver}.}
  \label{fig:observations_lc}
\end{figure}

\subsection{SN 2024ggi discovery and the multiwavelength observations}
SN 2024ggi was discovered by the Asteroid Terrestrial-impact Last Alert System (ATLAS) \cite{Tonry_2018} on 11 April 2024 03:23 UT (MJD 60411.14) in the galaxy NGC 3621 at a distance of 7.24~Mpc (Cepheid-based distance \cite{Saha_2006}, as used in \cite{Jacobson-Galán_2024, Shrestha:2024jal}). Soon after, it was classified as SN II with flash features, being clear signatures of photons from the embedded shock interacting with the CSM around it \cite{Zhang_2024}. The mass loss rate of the progenitor before the SN explosion lies in the range 10$^{-3}$--10$^{-2}$ M$_{\odot}$yr$^{-1}$, and the CSM velocity has been estimated at about 37 km s$^{-1}$ \cite{Shrestha:2024jal} (the values adopted in \ref{subsec:models}). SN 2024ggi has been the subject of extensive multiwavelength monitoring, owing to its proximity and early discovery. Besides multiple optical observations, X-ray detection was reported by NuSTAR \cite{nustar_atel} and EP-FXT \cite{xray_atel}, radio detection by ATCA \cite{radio_atel}, centimeter-wavelength upper limits were established by the JVLA and the uGMRT \cite{cm_atel}, and gamma-ray (100~MeV~--~500~GeV) flux upper limits by Fermi-LAT \cite{fermi_atel}.

\subsection{H.E.S.S. observations and data analysis}
The H.E.S.S. Target of Opportunity observations were carried out mostly with the full array of five telescopes, starting already on the same day the discovery was reported. This allowed for probing the system's early evolution during the rapid rise of the optical lightcurve, before it flattened out. Figure~\ref{fig:observations_lc} shows the total $\sim$~30 hours of observations after the data selection (excluding the data affected by bad atmospheric conditions and technical issues) spread over a month. The optical lightcurve based on the data provided by 
 the ATLAS forced photometry server \cite{Tonry_2018,atlas_transients,forcedphotoserver} is also shown in either of two filters, cyan (a bandpass between 420 and 650 nm) or orange (between 560 and 820 nm). A longer break between the observations was primarily caused by poor visibility due to a high level of moonlight (with a complete pause 23--25 April around the full Moon).
 
 The acquired data from 12-m telescopes (CT1--4) were analyzed using the ImPACT reconstruction framework \cite{impact} and cross-checked with an independent analysis chain \cite{modelplusplus}. The data from the large central telescope (CT5) were also examined \cite{Murach:2016SC}, accessing lower energy threshold. The high-level analysis was performed using the open source Gammapy software package (version 1.1) \cite{gammapy}.
The sky maps (excess and significance) were produced around the source position, in search for the potential gamma-ray signal, applying the ring background method \cite{bkg}. The significance was estimated following the approach in \cite{lima}. Furthermore, the differential photon flux upper limits were derived at the 95\% confidence level, using the reflected background method \cite{bkg}, and assuming a power law spectrum (dN/dE $\propto$ E$^{-\Gamma}$) with a photon index $\Gamma$ = 2. Similarly, the energy flux was calculated (for CT1--4 in the range 0.15--10~TeV) and converted to luminosity in order to compare with theoretical predictions.

\begin{figure}[b]
  \centering
 \begin{subfigure}[t]{0.33\textwidth}
    \centering
    \includegraphics[width=\linewidth]{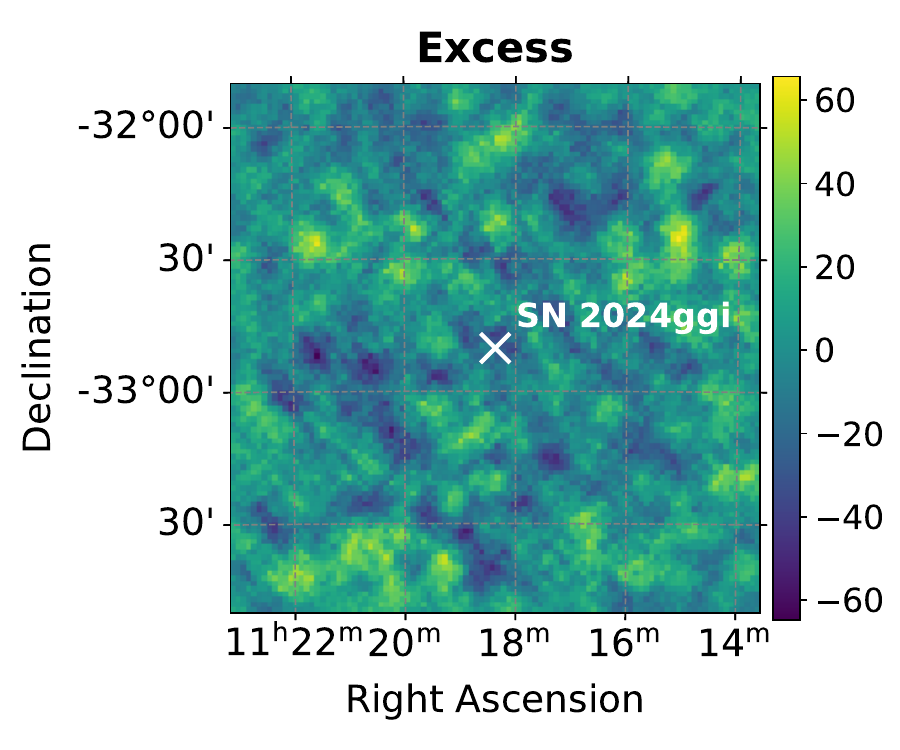}
    %\caption{Excess counts map}
    \label{fig:excess_map}
  \end{subfigure}
  \hspace{-2mm}
  \begin{subfigure}[t]{0.33\textwidth}
    \centering
    \includegraphics[width=\linewidth]{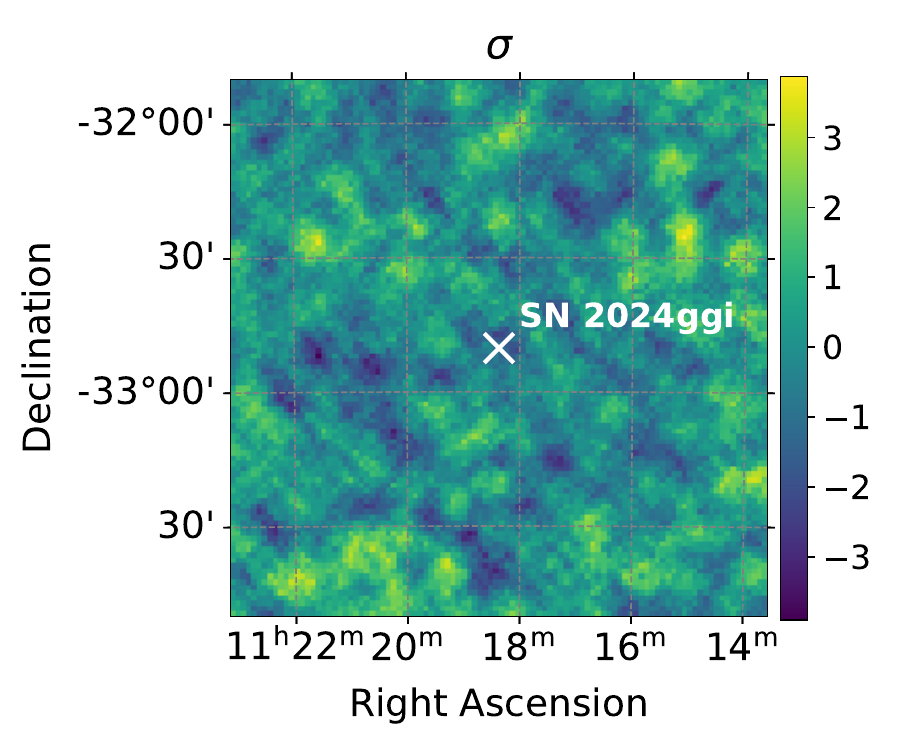}
   % \caption{Significance map}
    \label{fig:significance_map}
  \end{subfigure}
  \hspace{-2mm}
 \begin{subfigure}[t]{0.315\textwidth}
    \centering
    \includegraphics[width=\linewidth]{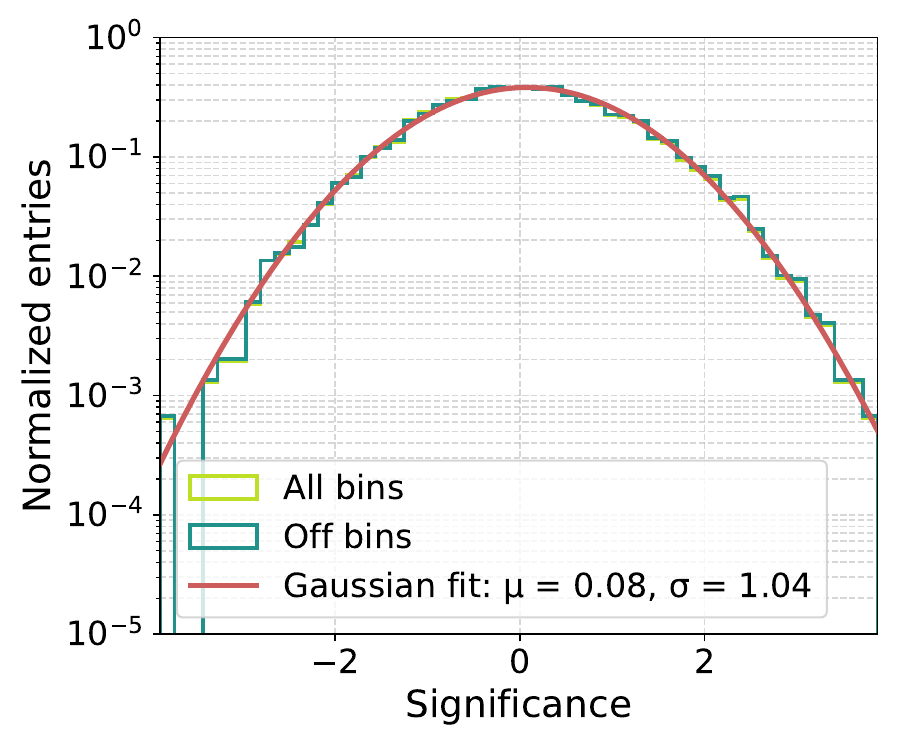}
   % \caption{Significance map}
    \label{fig:histogram}
  \end{subfigure}
  
  \caption{Excess count map (left panel), significance map (middle panel), and significance distribution with a Gaussian fit (right panel), computed from all the selected H.E.S.S. CT1--4 observations of SN 2024ggi.}
  \label{fig:maps_combined}
\end{figure}

\section{Results}
\begin{figure}[t]
  \centering
  \includegraphics[width=0.61\textwidth]{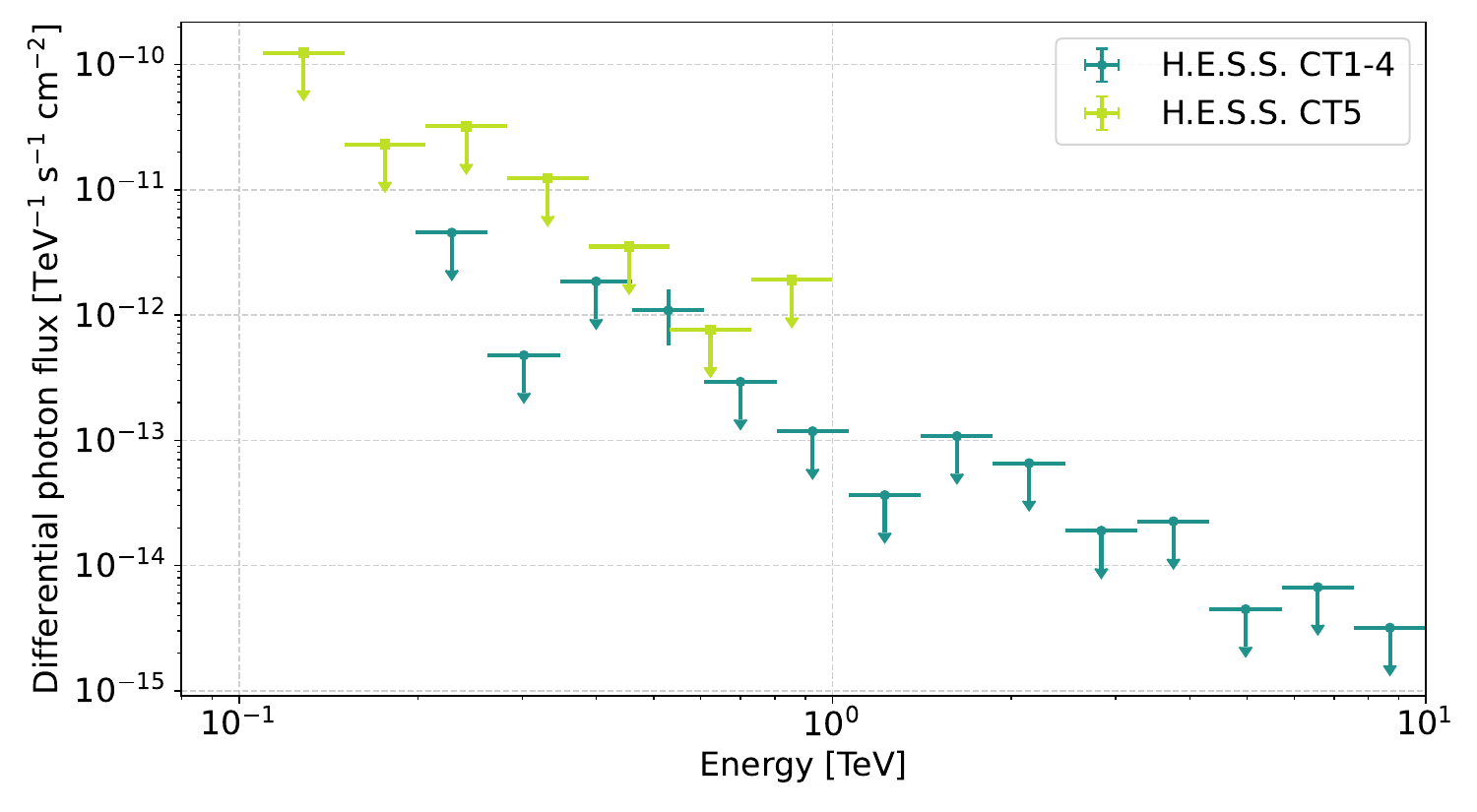}
  \caption{95\% C. L. differential photon flux upper limits, derived from all the selected H.E.S.S. observations of SN 2024ggi. Central telescope results (CT5) are shown in light green, and the results from the CT1--4 array are shown in dark green.}
  \label{fig:flux}
\end{figure}

\begin{figure}[b]
  \centering
  \includegraphics[width=0.61\textwidth]{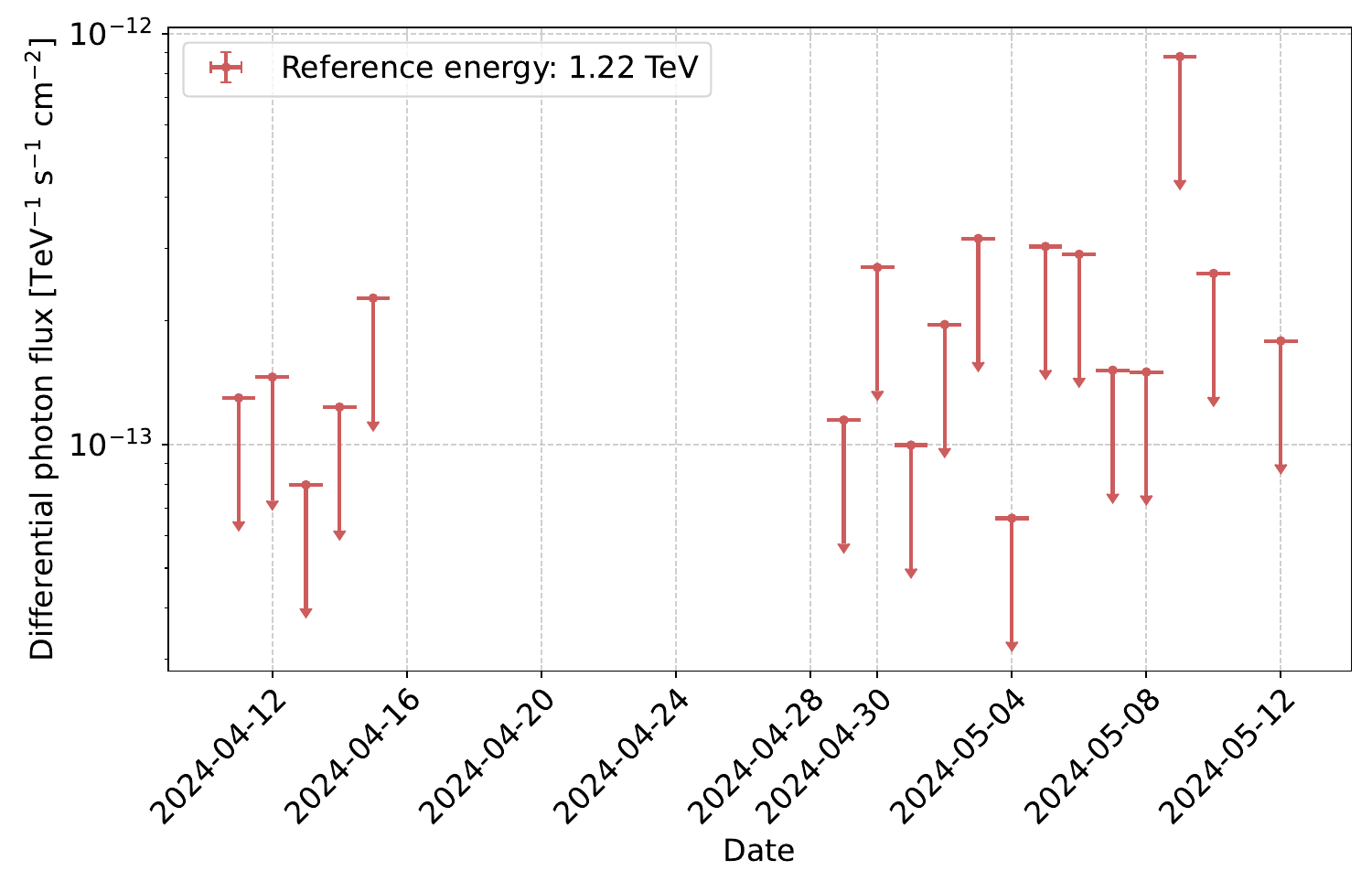}
  \caption{95\% C. L. differential photon flux upper limits at a reference energy of 1.22 TeV across the whole time of H.E.S.S. SN 2024ggi observations by the CT1--4 telescope array.}
  \label{fig:lc}
\end{figure}

The excess count map derived from all the selected CT1--4 H.E.S.S. observations of the SN 2024ggi is shown in Figure~\ref{fig:maps_combined}, together with the significance map and its distribution fitted with a Gaussian (where the mean, $\mu$ = 0.08, and the standard deviation, $\sigma$ = 1.04). No significant emission has been found at the source position (both in the CT1--4 and CT5 datasets, so we decide to show results only for the 12-m telescopes). 
Furthermore, we derive the 95\% C. L. differential photon flux upper limits, as shown in Figure~\ref{fig:flux} (also for large central telescope, CT5, reaching lower energies). We also show the upper limits across time of observation (with 1-day binning) at a reference energy of 1.22 TeV (midpoint on the logarithmic scale of energy range) in Figure~\ref{fig:lc} (only for CT1--4, as the CT5 data provide weaker constraints). 

Besides using the combined datasets, we also split the observations, focusing on the period with the lowest predicted level of the signal suppression due to $\gamma\gamma$ absorption ($<$ 1 day and $>$ 10 days after the SN explosion \cite{Cristofari_2020}). This yielded similar results, with no detection. 

\subsection{Comparison with theoretical predictions}
\label{subsec:models}
\begin{figure}[t]
  \centering
  \includegraphics[width=0.63\textwidth]{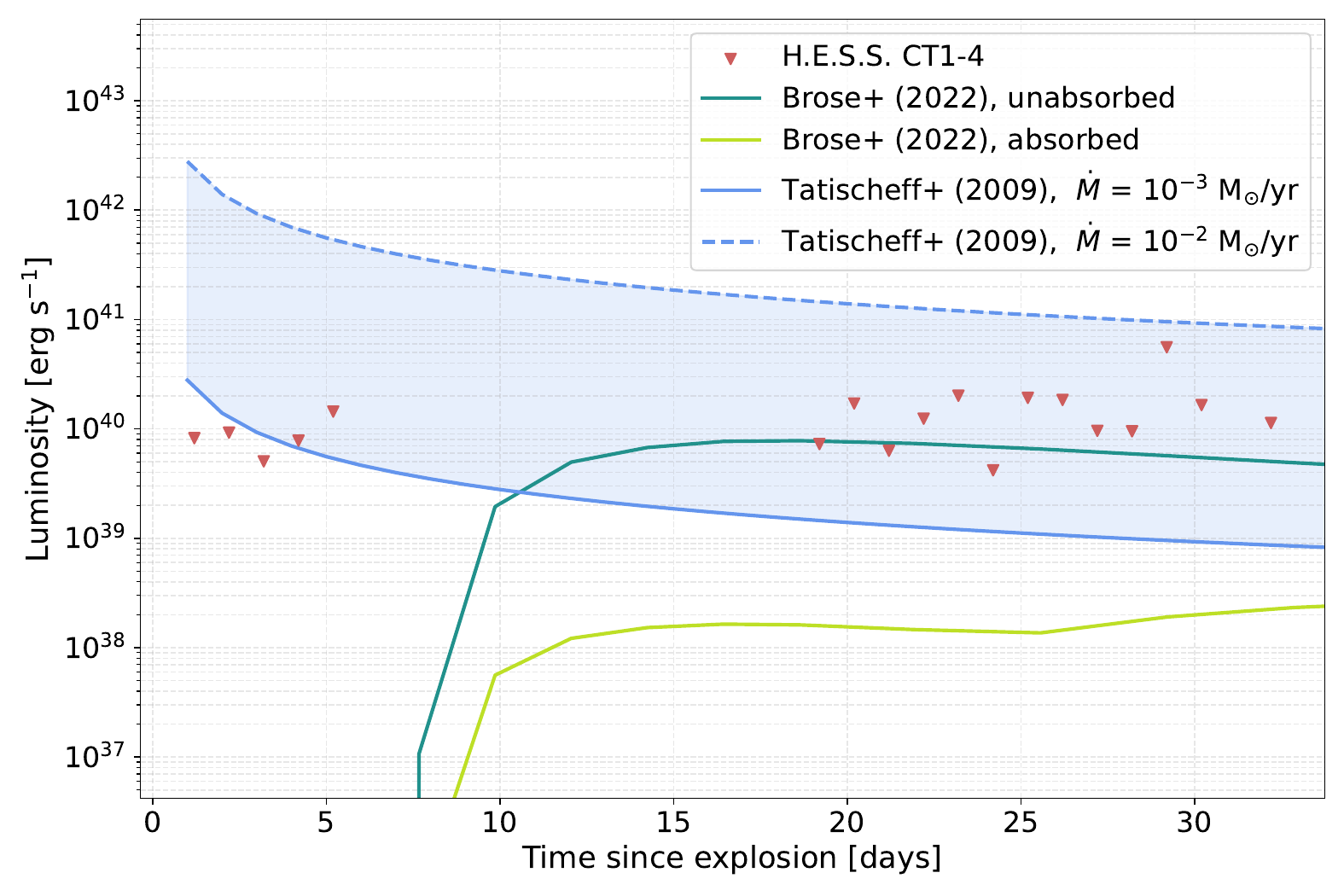}
  \caption{Comparison of the luminosity upper limits derived from H.E.S.S. CT1--4 observations of SN~2024ggi to the theoretical model from \cite{Brose_2022} with (light green) and without (dark green) the effect of $\gamma \gamma$~absorption included, as well as the analytical model from \cite{Tatischeff_2009}, applied to a possible range of mass-loss rate values (blue). }
  \label{fig:model}
\end{figure}

The energy flux upper limits have been computed for all the selected H.E.S.S. observations of SN 2024ggi and converted to luminosity, given the distance to the source of 7.24~Mpc. We present the results across time since the explosion epoch (MJD 60410.8 \cite{Shrestha:2024jal}) for CT1--4 only, since the CT5 data impose weaker limits. The comparison with theoretical predictions is shown in Figure~\ref{fig:model}. We include the model from \cite{Brose_2022}, which calculates the gamma-ray emission in the 1--10~TeV range and implements the attenuation of the signal due to the $\gamma \gamma$ absorption effect. We also show the analytical model from \cite{Tatischeff_2009}, following \cite{Cristofari_2020} as in \cite{Mart_Devesa_2024}, Equation~4. The values of the parameters related to the system, such as distance, shock velocity, wind velocity, and mass-loss rate are taken from \cite{Shrestha:2024jal} (which are similar to other estimates found in the literature \cite{Zhang_2024, Chen:2024mye, Jacobson-Galán_2024}). Our upper limits on day 1 following the explosion are inconsistent with the bright early gamma-ray emission predicted by \cite{Tatischeff_2009, Cristofari_2020}. Later upper limits are consistent with predictions using wind densities estimated from optical observations.

\section{Conclusions}
The H.E.S.S. observations of a nearby SN 2024ggi, conducted as a part of the core-collapse SN follow-up program,  were presented. While no significant signal excess was found, the very-high-energy upper limits provide important insights to the overall multiwavelength picture of such sources. Moreover, the measurements can be used to constrain parameters of the system, offering meaningful information for theoretical predictions, also regarding the acceleration of cosmic rays and gamma-ray detectability (which is influenced, for instance, by the distance or the level of $\gamma \gamma$ absorption).

% \newpage

\printbibliography

\section*{Acknowledgements}

{\footnotesize

The support of the Namibian authorities and of the University of Namibia in facilitating the construction and operation of H.E.S.S. is gratefully acknowledged, as is the support by the German Ministry for Education and Research (BMBF), the Max Planck Society, the Helmholtz Association, the French Ministry of Higher Education, Research and Innovation, the Centre National de la Recherche Scientifique (CNRS/IN2P3 and CNRS/INSU), the Commissariat à l’énergie atomique et aux énergies alternatives (CEA), the U.K. Science and Technology Facilities Council (STFC), the Polish Ministry of Education and Science, agreement no. 2021/WK/06, the South African Department of Science and Innovation and National Research Foundation, the University of Namibia, the National Commission on Research, Science \& Technology of Namibia (NCRST), the Austrian Federal Ministry of Education, Science and Research and the Austrian Science Fund (FWF), the Australian Research Council (ARC), the Japan Society for the Promotion of Science, the University of Amsterdam and the Science Committee of Armenia grant 21AG-1C085. We appreciate the excellent work of the technical support staff in Berlin, Zeuthen, Heidelberg, Palaiseau, Paris, Saclay, Tübingen and in Namibia in the construction and operation of the equipment. This work benefited from services provided by the H.E.S.S. Virtual Organisation, supported by the national resource providers of the EGI Federation.}

{\footnotesize
This work has made use of data from the Asteroid Terrestrial-impact Last Alert System (ATLAS) project. The Asteroid Terrestrial-impact Last Alert System (ATLAS) project is primarily funded to search for near earth asteroids through NASA grants NN12AR55G, 80NSSC18K0284, and 80NSSC18K1575; byproducts of the NEO search include images and catalogs from the survey area. This work was partially funded by Kepler/K2 grant J1944/80NSSC19K0112 and HST GO-15889, and STFC grants ST/T000198/1 and ST/S006109/1. The ATLAS science products have been made possible through the contributions of the University of Hawaii Institute for Astronomy, the Queen’s University Belfast, the Space Telescope Science Institute, the South African Astronomical Observatory, and The Millennium Institute of Astrophysics (MAS), Chile.
}

\end{document}